\RequirePackage[2020-02-02]{latexrelease}
\documentclass[aps,preprint,a4paper]{revtex4}%
\usepackage{amsfonts}
\usepackage{amsmath}
\usepackage{graphicx}
\usepackage{amssymb}%
\providecommand{\U}[1]{\protect\rule{.1in}{.1in}}

\ifx\pdfoutput\relax\let\pdfoutput=\undefined\fi
\newcount\msipdfoutput
\ifx\pdfoutput\undefined\else
\ifcase\pdfoutput\else
\ifx\paperwidth\undefined\else
\ifdim\paperheight=0pt\relax\else\pdfpageheight\paperheight\fi
\ifdim\paperwidth=0pt\relax\else\pdfpagewidth\paperwidth\fi
\fi\fi\fi
\begin{document}
\title{Quantum Time: a novel resource for quantum information}
\author{M. Basil Altaie}
\affiliation{The School of Physics and Astronomy, University of Leeds, Leeds LS2 9JT,
United Kingdom. }
\pacs{PACS number}

\begin{abstract}
Time in relativity theory has a status different from that adopted by standard
quantum mechanics, where time is considered as a parameter measured with
reference to an external absolute Newtonian frame. This status strongly
restricts its role in the dynamics of systems and hinders any formulation to
merge quantum mechanics with general relativity, specifically when considering
quantum gravity. To overcome those limitations, several authors tried to
construct an operator which is conjugate to the Hamiltonian of quantum systems
implementing some essential features of the relativistic time. These
formulations use the concept of internal or intrinsic time instead of the
universal coordinate time used in textbooks. Furthermore, recently it is
remarked that the consideration of time with relativistic features could
enhance the analysis techniques in quantum information processing and have an
impact on its status in causal orders and causal structures of quantum
information. The role of clocks, their accuracy and stability has become an
important issue in quantum information processing. This article present a
substantiative review of recent works which reflect the possibility of
utilizing quantum time, measured by quantum clock devised according to
Page-Wootters scheme, to stand as a resource for quantum information processing.

\end{abstract}
\startpage{1}
\endpage{102}
\maketitle

\section{Introduction}

Time is enigmatic, illusive and is profoundly connected with our
consciousness. Physically, it is a measure of change, intrinsically associated
with the dynamics of physical systems. The nature, the role, and the meaning
of time has been contemplated and discussed by philosophers and scientists
throughout the ages. Isaac Newton comprehended time as a reality of existence
driving the dynamics of change \cite{New}, while Albert Einstein finds that
time is a relative measure that depends on the frame of reference, and in some
respect is an illusion \cite{Bar}. Time signifies our own existence as
conscious beings despite our inability to realize time in its objective
reality as we do with space, motion and matter.

\begin{figure}
[h]
\begin{center}
\includegraphics[
width = 0.43\textwidth
]{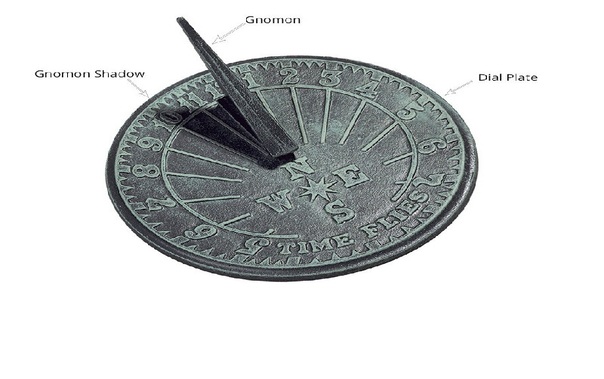}
\caption{Sundial marks the local solar time. This is an intrinsic natural time
that depends on the location of the sundial. }%
\label{F1}%
\end{center}
\end{figure}

There are several methods for measuring time, all are meant to mark the
occurrence of events orderly. The timer used in lab is treated as an external
entity that has nothing to do with the system under measurement. This is
called the \textit{coordinate time}. Sundials measure the local solar time as
marked by the shadow of an object (the gnomon) accounted for by the apparent
movement of the Sun in the sky. Unlike our wristwatch a sundial reads an
intrinsic time which is local. The readings of the shadow of the gnomon (see
Fig. \ref{F1}) are entangled with the apparent positions of the Sun in the
sky, these readings are different from those on your wristwatch. This is
because your wristwatch does not tell you the \textit{natural time} but the
\textit{mean solar time}, which is conventional. Time in your wristwatch is
running at a constant rate, but the time readings of a sundial does not run at
a constant rate. However, we can always use any of these systems or any other
periodic system to act as a clock. This is an indication that time is not an
absolute entity but a measure of change. It is a `relatival' entity that
acquires its meaning through the relational dynamics of change.%

In the Newtonian realization, time is an entity that exist independent of
space, motion and matter \cite{New}. It is like a continuous flow that runs
equally through the world. In the Newtonian mechanics time is an essential
external parameter in the dynamics of the world.

\begin{figure}
[h]
\begin{center}
\includegraphics[
width = 0.38\textwidth]{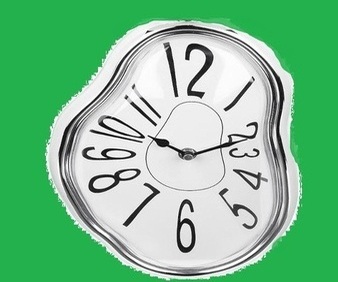}
\caption{Dali-type clock with uneven divisions showing dilated times as it is
the case in Sundials.}
\label{F2}
\end{center}
\end{figure}

The theory of relativity recognized that space and time form one integrated
covariant entity, called \textit{spacetime}, and that both depend on the frame
of reference of the observer. Time is found to dilate with speeding and also
in places of strong gravity. In this sense time becomes an intrinsic property
that depend on the frame of reference. This is like the time measured by a
Dali's clock, after the melting clocks of Salvador Dali, Fig. (\ref{F2}). It
is here where the artistic prefiguration get verified by the physical theory.
The theory of general relativity teaches us that there is nothing outside the
universe, the universe is a block with space and time forming a continuum
preserving the covariance of the spacetime interval everywhere \cite{Wein}.
Space is not a preset stage for events and time is not an absolute measure of
change, both are relative and observer dependent. This may suggest that the
identification of the dynamics in space and time become \textit{relational}.%

Standard textbook quantum mechanics\ adopts the Newtonian concept of time.
Time is an independent continuous parameter, measured by an external clock
which has its reference with an absolute hypothetical time. In spite of being
an essential factor in quantum dynamics, time is not a dynamical observable
\cite{Per}, no Hermitian conjugate operator of time is known. Historically,
this stand was strongly influenced by Pauli's objection \cite{Pau} arguing
that unless the Hamiltonian is unbound from below, then both time and energy
states cannot establish correlations. However, adopting a new formalism for
the measurement of time by means of quantum clock could avoid the Pauli
objection \cite{Leon}.

Quantum time and the use of quantum clocks has been discussed by several
authors and several proposals are put forward. Salecker and Wigner \cite{SW}
introduced the idea of a quantum clock where they propose to use clocks only
for measuring spacetime distances and avoid using measuring rods, which are
essentially macro-physical objects. Connes and Rovelli \cite{CR} tried to
relate problems connected with the notion of time in gravitation, quantum
theory and thermodynamics using the algebraic formulation of quantum theory
thus proposing a unifying perspective of those problems. In their proposal the
physical time-flow appears to be determined by the thermodynamical state of
the system. More recently Pearson et al., \cite{Person} verified the
relationship between the time-flow in a clock and the thermodynamics and found
a linear relationship between the accuracy of the clock and entropy where the
quality of the clock gets better as the entropy increases. All these proposal
are interesting and worth discussing. However, the intriguing suggestion came
by the formulation proposed by Page and Wootters (PaW) \cite{PW} in which they
established the conceptual and mathematical structure of a novel construction.
The main point is to adopt the notion of internal quantum time represented as
an operator conjugate to the Hamiltonian in contrast to the external
coordinate time which is adopted as a \ parameter in the standard formulation
of quantum mechanics. This was a step forward toward merging relativity with
quantum mechanics. The second major contribution in this respect is the
adoption of the constraint Hamiltonian composed of the Hamiltonians for the
quantum clock and the rest acting on a global wavefunction for the whole
system (called the universe). In the light of the development and renovations
of the PaW scheme that took place during the last few years and the growing
interest in this scheme we found the need to present some suggestions in
respect of employing quantum time in quantum information processing.

In this article we discuss the implication of quantum time through some
applications in gravity and quantum information. The detailed mathematical
formulations of quantum time according to the PaW scheme and the construction
of quantum clocks can be found in \cite{Altaie}. So, there is no need to
repeat the formulations here except whenever necessary upon constructing the
clocks for the specific problems we are dealing with. In Sec. (2) we elaborate
more on the potential relevance of quantum time for solving outstanding
problems in quantum measurements. In Sec. (3) we give an overview of the main
content of the quantum time formulation and description of the quantum clock
including the peculiar way in which the progression of time is viewed. In Sec.
(4) the quantum time is introduced into quantum information processing, where
we expose two important features of it, namely its dependence on the frame of
reference where the need for a quantum `proper time' appears, and its bearing
on the causal order of events and causal structures. Finally in Sec. (5) we
summarize the main aspects of quantum time setting the available techniques to
the scrutiny of experimental verification and pointing to possible future
venues of potential research.

\section{Outstanding Problems}

Most problems in quantum dynamics which involve measurement of time are in
essence quantum information problems. There are several outstanding problems
in quantum mechanics that has found no convincing explanations.

In addition to the infamous paradoxes of Schr\"{o}dinger's cat and the EPR,
there are several other unresolved problems involving time in conventional
quantum applications, this includes the speed of quantum tunneling in what is
known as the Hartmann effect \cite{Hart}. Such an effect has been verified
experimentally by many experiments using electrons \cite{Sek} and photons
\cite{Stein}. Some experiments suggest that the penetration takes no time at
all \cite{Sain}, others are suggesting a superluminal speed \cite{EN,EN1,Heit}%
. Several explanations have been given for these experiments \cite{Win1,Ramos}%
, but the most interesting ones are those which involve calculating the
arrival time with a quantum clock \cite{MS}.

The quantum Zeno effect \cite{Sud} is another fundamental problem in quantum
mechanics in need of being presented and analyzed in the context of quantum
information (for a thorough account of the quantum Zeno effect see for example
\cite{Pao}). Again here the concept and the measurement of time plays an
important role in assessing the phenomenon. Peres \cite{Pe1,Per1} discussed
this in the context of his formulation of the quantum measurement of time.
However, Beige and Hegerfeldt \cite{Beige} have shown analytically that, for a
wide range of parameters, the usual projection postulate applies with high
accuracy. Currently it is known that the quantum Zeno effect can be used to
protect appropriately encoded arbitrary states to arbitrary accuracy, while at
the same time allowing for universal quantum computation or quantum control
\cite{Beige1,Paz}. These results has already initiated a great deal of works
on the temporal optimization of quantum computing (see \cite{Yig,An} and the
references therein).

The question of arrival time in quantum mechanics is one more topic which is
related to the measurement of time. Within the recent formulation of quantum
time measurements Maccone and Sacha \cite{MS} were able to present a general
and successful formulation for measuring the arrival time. More recently
Gambini and Pullin \cite{GP} showed that the quantum time formulation provides
a natural solution to the problem of the time of arrival and leads to a well
defined time-energy uncertainty relation for the clocks. This confirms the
role of quantum time calculation in solving problems faced by standard quantum
information theory. This is a serious information-theoretic achievement
stressing the role of quantum time.

One of the topics where we face a serious problem with time is quantum gravity
\cite{Kuc,Isam,Asht,Andr}. The canonical quantization of gravity, as described
by the theory of general relativity, inevitably yields the Wheeler-DeWitt
equation \cite{Dew,HH}. In one form this equation describes a Hamiltonian
constraint, with a timeless (static) block universe as viewed by an external
hypothetical observer. For an observer inside the universe physical systems
are under temporal development driven by the dynamics ruling the system. In
one facet of the problem, having time expressed by an operator is necessary
to\ quantize gravity, an essential step to understand the origin of the
universe and the fate of black holes. For example, resolving the information
paradox in black hole physics \cite{Haw} and understanding what happens inside
the event horizon requires a quantum description of gravity \cite{Raj}.
Furthermore resolving the information paradox is necessary to preserve the
unitarity of physics. The same theory is needed to understand the very early
moments of the development of the universe and the role played by gravity in
getting something out of nothing \cite{Kraus}.

In formulating quantum gravity, many conceptual issues turn out to be related
to the formulation of time. This occurs because gravity is essentially a
curvature of time. The leading term in the time-time component of the metric
tensor in the external Schwarzschild solution $g_{00}=\left(  1-2GM/c^{2}%
r\right)  $ is proportional to the gravitational potential. Several approaches
to formulate a viable theory of quantum gravity fell short of achieving the
goal and others are still surviving. In all cases time is at the heart of the
problem \cite{Rov1,Rov2,Smolin,Bar}. This shows how serious it is to consider
the measurement of time as a basic information-theoretic problem.

\section{The Quantum Time}

Associating the word quantum with anything often implies that it is discrete,
but certainly it is not limited to that. Having time as a parameter does not
qualify it to be a generator of dynamics. Furthermore, having the time
measured in respect to an independent external reference does not provide us
with any sort of relational internal information system. Within a closed
system time have its status as an internal measure which would qualify it to
play the role of a connection signifying the order in a relational dynamics of
events. To achieve these two requirements we have to treat time as an operator
conjugate to the Hamiltonian of the quantum system and have it measured by an
internal quantum clock, which is another quantum system, for example the
precision of a spinning elementary particle can be considered as a clock
correlated with the transitions of the quantum states \cite{Per1}. The
question is: How can such a construction, which take us beyond the
conventional formulation of quantum mechanics, be formulated? As we are
interested in presenting the implications and applications of quantum time in
quantum information processing it is necessary to realize the connection
between the structure of quantum time, being an internal measure, and the
confined character of the quantum information processing. For this we need at
least a brief description of the basic mechanism of quantum time, more details
on this can be found in \cite{Altaie} and the references therein.

\subsection{The Page and Wootters formalism}

Page and Wootters based their formulation on several fundamental arguments
\cite{PW}. The first is that any quantum observable must be stationary, which
means that it has to commute with the Hamiltonian. The second is that the
calculated probabilities in quantum mechanics are to be considered
conditional, meaning that the state $|\psi(t)\rangle$\ is conditioned to exist
at the time $t$. The third assumption of PaW is that the possible states of a
quantum system all form one global time-independent state $|\Psi\rangle$ that
satisfies the constraint%
\begin{equation}
H|\Psi\rangle=0, \label{q2}%
\end{equation}
where $H=H_{c}+H_{s}$with $H_{c}$ being the clock Hamiltonian and $\ H_{s}$ is
the system Hamiltonian. The Hamiltonian in Eqn. (\ref{q2}) is described in a
joint Hilbert space which is a direct product of the two Hilbert spaces
$\mathcal{H}_{c}\otimes\mathcal{H}_{s}$ where $\mathcal{H}_{c}$ is spanned by
the basis states of the clock having dimension $d_{c}$ and $\mathcal{H}_{s}$
is spanned by the basis states of the quantum system and have a dimension
$d_{p}$. The dimension of the clock's Hilbert space must be larger than the
dimension of the system's Hilbert space in order for the clock to cover all
possible states of the system. If the clock Hamiltonian is taken as
$H_{c}=-i\hbar\frac{\partial}{\partial t_{c}}$ then Eqn. (\ref{q2}) reads%

\begin{equation}
\left(  H_{s}-i\hbar\frac{\partial}{\partial t_{c}}\right)  |\Psi\rangle=0.
\label{q2a}%
\end{equation}
The Schr\"{o}dinger state at a time $t$ can be obtained by projecting
$|\Psi\rangle$ onto $t$. So,%

\begin{equation}
|\psi(t)\rangle=\langle t|\Psi\rangle. \label{q4}%
\end{equation}
This is the basic structure of the formulation used in recent works presenting
the quantum time measurement scheme. The original PaW formulation was
criticized by Kuchar \cite{Kuch}, and \cite{Unr} for a number of vital points,
mainly concerning the progress of the clock and some other ambiguities.
Gambini \cite{Gam} and later \cite{MV} have set proposals to resolve those
questions and ambiguities of the original PaW scheme using Rovelli's `evolving
constants'. Eventually the PaW formalism was renovated by Giovannetti and
collaborators \cite{G}, Marletto and Vedral \cite{MV} and Maccone and Sacha
\cite{MS}. .

\subsection{The quantum clock}

As mentioned above there has been different choices of the clock, however all
are basically quantum systems with well-defined energy spectrum defining the
basic structure of the clock Hamiltonian. But it worth mentioning that the
approach in \cite{FS, Boe1,Boe2} adopting a prescription proposed by Pegg
\cite{Pegg}, in which the clock has a finite number of states with a
Hamiltonian bounded from below, might be more suitable to use for dealing with
quantum information problems. Nevertheless the mathematical machine is the
same. The clock Hamiltonian $H_{c}$ is described in a Hilbert space
$\mathcal{H}_{c}$\ of finite or infinite dimensions. The bedrock in this
construction is to define a Hermitian time operator conjugate to the
Hamiltonian. In this case $H_{c}$ is shown to be a generator of shifts in the
readings of the clock and vice-versa the time operator is a generator of
energy shift (see for example \cite{FS}), The readings of the clock (time
states of the clock) are entangled with the possible observables described by
the states of the system. This is achieved by applying the constraint as above
in Eqn. (\ref{q2a}). But in order to maintain independence of the clock we
should maintain that it does not interact with the quantum system \cite{PW}.
This is done by having the time operator commuting with the Hamiltonian,
consequently the states of the clock are stationary. 

\subsection{Dynamics}

The global state $|\Psi\rangle$ is a solution of (\ref{q2a}). For a finite
dimensional clock space $d_{s}$ is given by \cite{FS}
\begin{equation}
|\Psi\rangle=\frac{1}{\sqrt{d_{s}}}\sum\limits_{m=0}^{s}|\tau_{m}\rangle
_{c}\otimes|\phi_{m}\rangle_{s}, \label{q11}%
\end{equation}
where $|\phi_{m}\rangle_{s}$ is a generic state of the quantum system given
by\qquad\
\begin{equation}
|\phi_{m}\rangle_{s}=\sum\limits_{n=0}^{p}c_{k}e^{-iE_{k}\tau_{m}/\hbar}%
|E_{k}\rangle. \label{q12}%
\end{equation}
In this case $H_{c}$ is shown to be a generator of shifts in the readings of
the clock and vice-versa the time operator is a generator of energy shift
given by%

\begin{equation}
|\tau_{m}\rangle_{c}=e^{-i\overset{\wedge}{H}(\tau_{m}-t_{0})/\hbar}|\tau
_{0}\rangle_{c}, \label{Eq7}%
\end{equation}
and%

\begin{equation}
|E_{n}\rangle_{c}=e^{i\overset{\wedge}{\tau}(E_{n}-E_{0})/\hbar}|E_{0}%
\rangle_{c}. \label{q8}%
\end{equation}
From Eq. (\ref{q11}) we can easily see that
\begin{equation}
|\phi_{m}\rangle_{s}=d_{c}\langle t_{m}|\Psi\rangle. \label{q13}%
\end{equation}
Now employing this in the PW mechanism using Eq. (\ref{q2a}) and Eq.
(\ref{Eq7}) the states develops in quantum time as
\[
|\phi_{m}\rangle_{s}=e^{-i\overset{\wedge}{H}_{s}(\tau_{m}-\tau_{0})/\hbar
}|\phi_{0}\rangle,
\]
this is the Schr\"{o}dinger equation for the time-development of the states of
the quantum system correlated with the time measured by a quantum clock. The
conditional probability of obtaining an outcome $a$ for the system $S$ when
measuring the observable $A$ at a certain clock time $\tau_{m}$ \ is expressed
according to the Born rule as
\begin{equation}
P\left(  a\text{ on }S|\tau_{m}\text{ on }C\right)  =|\langle a|\overset
{\wedge}{U}_{s}(\tau_{m}-\tau_{0})|\phi_{0}\rangle|^{2}. \label{qJ}%
\end{equation}
where $\overset{\wedge}{U}_{s}(\tau_{m}-\tau_{0})$ is the transition matrix
between the clock states correlated with the eigen states of the quantum system.

\subsection{Progression of time}

\begin{figure}
[ptb]
\begin{center}
\includegraphics[
width = 0.72\textwidth
]{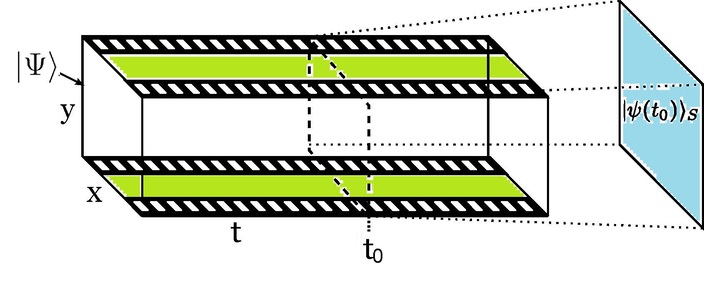}
\caption{States of the quantum system as seen by an internal observer,
time-developing according to Schr\"{o}dinger's equation. The global
wavefunction is depicted as a reel of frames of a movie with respect to an
external observers.}%
\label{F3a}%
\end{center}
\end{figure}

Basically, the global system which includes the clock and the quantum system
form a `block universe'. Events (quantum states) are stationary, so
effectively there is no flow of time with respect to a hypothetical external
observer. However, an observer within this universe will see a change of the
states as they get projected sequentially through the correlation of time
measurements and all the available states are contained in $|\Psi\rangle.$
These are the Schr\"{o}dinger wave functions $|\phi_{m}\rangle$ projected
every time an observation of a value of the observable is correlates with the
time $\tau_{m}$. The dynamics in this universe emerges as the clock advances
and the states of the quantum system get exposed in a measurement. This could
be understood in analogy with the sequence of movie snap shots (frames) stored
in a film reel, as shown in Fig. (\ref{F3a}). In this analogy, the shots of a
flash lamp corresponds to the ticks of the clock.%

\section{Simple examples}

There are some applications where the PaW scheme has been applied to physical
problems within this context. Several authors cited above considered such
applications in physical problems in spacetime, gravity and thermodynamic,
however, there seems to be a simpler approach that is followed by Favalli and
Smerzi which exposes several aspects of this scheme reflecting its richness
and clarity. Such works will be useful in providing the material for preparing
algorithms for information processing of such problems sooner or later.

In \cite{FS1} we find that when the PaW scheme is generalized by considering
two constraint one on the energy and the second on the linear momentum, a
$3+1$ dimensional model can be constructed and non-relativistic spacetime is
found to emerge from the entanglement of subsystems in a globally timeless and
positionless Universe. This work certainly needs more development in a wider
and more profound consideration. However, in the present formulation it may
serve to provide the basic formalism to study the relativistic case. In
\cite{FS2} the same authors present a thermodynamical treatment of a quantum
universe composed of a small system in a large environment. Again using the
energy constraint and the PaW mechanism they show that time and
non-equilibrium dynamics can emerge as a consequence of the entanglement
between the system and the environment (represented by the ticks of the clock)
described within the global wavefunction, thus concluding a peaceful
coexistence of thermal equilibrium and the emergence of time. This we find is
less complicated presentation of similar situation that has been discussed in
some respect by \cite{CS}. In \cite{FS3} two non-relativistic quantum clocks
interacting with a Newtonian gravitational field has been considered. The
authors derive a time dilation for the time states of the clocks. The delay is
in agreement up to first order with the gravitational time dilation obtained
from the Schwarzschild metric. Once extended by considering the relativistic
gravitational potential the result is in agreement with the exact
Schwarzschild solution.

The above examples show that the use of the constraint Hamiltonian acting on a
global wavefunction summarizes lots of information embedded in such a
formulation. The above examples show that a cluster of physical problems might
be connected within the same conception of a block universe and the same
mathematical formulation. This can be comprehended once we remember that
$H|\Psi\rangle=0$ is describing an entanglement between time (clock states)
and the states of the system confined in a closed universe.

\section{Prospects for Quantum Information Processing}

Quantum information processing deals with discrete time and is fundamentally
concerned with constructing\ a memory, a record information \cite{Gio}. The
existence of a memory is an essential part of information systems. It cares
for the causal order of events, without which no order can be identified.
Thus, partially ordered sets play a fundamental role in causality. The
\textit{before} and the \textit{after} are terms used to assign the causal
order of the events, and this trend of analyzing causal order \ and causal
structure has become an advanced trend of analysis \cite{Fey,Pearl1}. In this
section we briefly survey some known quantum information venues where quantum
time is expected to play a role, with an emphasis on a broad overview of their
applications rather than their technical details. The first issue is the
causal order which is part of the structural map of information processing and
the second is the importance of the quantum reference frame and its influence
in quantum information.

Manipulating information bits (qubits) back and forth as ordered states with
high sampling rate requires dealing with states within very short times
\cite{Lund}. The states being the qubits picked up, or recorded into a memory.
It is known that in such a process the challenge to the broad application of
quantum information sciences is the management of errors in a fragile quantum
hardware \cite{Nie}. Such quantum errors might involve temporal effects since
it is known that the prevalent component of decoherence of the superposition
of the states is dephasing which randomizes the relative phase between basis
states. Likewise. as the quantum states are correlated with the quantum clock,
the qubit coherence is inferred relative to a reference which is generally
taken to be a local oscillator used to interrogate the system \cite{Aud}. Ball
and collaborators\emph{ }\cite{Ball} have correctly remarked the importance of
master clock phase fluctuations as an emerging source of error in qubit
systems, and expected that this will become more prominent as the qubit error
rates continue to improve. Quantum algorithms are set for performing quantum
computations, and a resolution of the problems mentioned here through the
inspiration of advanced quantum algorithms, such as topological quantum field
theory \cite{Free}, quantum circuits and spin models \cite{Delas}\ has been
already sought during the last two decades as pointed to by Montenaro
\cite{Ash}. Hence it would be a natural development to introduce a quantum
time formalism into quantum algorithms in order to enhance the operational
fidelities by noise reduction. Indeed, very recently Diaz et al., \cite{Diaz}
presented a quantum algorithms for parallel-in-time simulations that are
inspired by the PaW formalism. They showed that their algorithms can compute
temporal properties over $N$ different times of many-body through using
$\log(N)$ only. Such a demonstration is certainly related to the equilibrium
of an isolated system, thus allowing for considering the dynamical properties
of many-body systems. described with intrinsic time formulation, under quantum
information processing.

The superposition of direct pure processes (SDPP) is attracting the attention
of many researchers \cite{Guer,Tadd,Krist,Rzep}. This is related to quantum
switches and causal order. An example is given by \cite{Guer}, where a qubit
is initially encoded in the polarization degrees of freedom of a photon, which
travels in an equal superposition of two paths. At the end, a localized
observer can simultaneously access both modes and perform a measurement to
recover information about the initial state of the polarization qubit. The
state of the photon at time $t$ (just before any noisy channel gets applied)
is the same in both implementations. This example reminds us of Wheeler's
delayed choice experiment \cite{Wheeler}, which is finding applications in
linear optics, nuclear magnetic resonance, and integrated photonic device
systems in the optical platform. Dong and collaborators \cite{Dong} have
demonstrated a Wheeler's delayed-choice experiment setup in an interface of
light and atomic memory, in which the cold atomic memory makes the heralded
single-photon divided into a superposition of atomic collective excitation and
leaked pulse, thus acting as memory-based beam-splitters. The results can be
interpreted to confirm Bohr's view that the wave-like or particle like
behavior of light and matter is exposed only when the measurement happens.
This would confirm again that the wave-like and time like properties of atoms
and light are complementarity features. A more recent article \cite{GS}
confirmed the same results for Rydberg atoms. Therefore, it would be of
interest to see the validity of introducing quantum time in the SDPP technique
to resolve the questions raised by Wheeler's thought experiment since it is
shown that such a process can be formulated as a quantum-controlled device
\cite{Ionic}, which involves quantum switching, a step which will extend the
applications of this phenomenon in quantum information processing. A modified
quantum delayed-choice experiment without quantum control or entanglement
assistance has been recently suggested \cite{Guo} in which a photon can be
prepared in a wave particle superposition state and the morphing behavior of
wave-to-particle transition can be observed easily. It is demonstrated that
the presented scheme can allow us to rule out the two-dimensional classical
hidden variable causal models in a device-independent manner through violating
dimension witness. The experiment is further extended to the situation of two
degrees of freedom, which enabled simultaneous observation of a photon's wave
and particle behaviors in different degrees of freedom, and then proposing a
scheme to prepare the single photon wave-particle entanglement. This last
remark by the authors is very important since it inspires one to think about
deducing the single-particle nonlocality from the perspective of the
wave-particle degree of freedom, which means that the morphing behaviour of
the wave-to-particle is resulting out of periodic change in the existence of
the particle getting reflected as a statistical wave-like behavior. On a more
basic level this has been already suggested long ago by de Broglie himself
\cite{de Broglie}. In searching for the de Broglie internal clock of the
particle an experiment was performed \cite{Cat} showing that the particle
state can itself be identified as an internal clock. This could provide an
intuitive picture to the principle of complementarity pointed to above once we
consider the state of the particle is under continued re-creation.

\subsection{Causal Order and Causal Structures}

A fixed background causal structure is traditionally assumed to exist in
quantum processes. However, if the laws of quantum mechanics are applied to
the causal relations, then one may expect that the causal order of events is
not always fixed since the quantum uncertainty would be at play. Such
indefinite causal structures could make new quantum information processing
tasks possible and provide methodological tools in quantum theories of gravity
\cite{Buck}. During the last decade or so there has been much interest in
investigating the question of causal order and causal structure in connection
with relational dynamics which seems to be requested by the unification of
quantum mechanics and relativity theory. Certainly such investigations are
necessary to provide a clear understanding of the information processing
themes in a quantum time framework under relational dynamics. An interesting
work \cite{Ore}\ was published ten years ago in which the authors tried to
investigate the question: where does causal order come from and whether it is
a necessary property of nature? This was addressed into a multipartite
correlations without assuming a pre-defined global causal structure except the
validity of quantum mechanics locally. The authors find that correlations
cannot be understood in terms of definite causal order. Such correlations,
they claim, violate a `causal inequality' which is satisfied by all space-like
and time-like correlations. On assessing the applicability of such a
conclusion we recognize that the requirement of the validity of quantum
mechanics locally implicitly allows for the availability of local hidden
variables to be at play. A more general consideration must do without such a
condition and this is what has been very recently exposed in \cite{Costa}
where the author analyze the problem more systematically, admitting that
causal order of events need not to be fixed primarily. In his words
\textquotedblleft whether a bus arrives before or after another at a certain
stop can depend on other variables, like traffic.\textquotedblright%
\ Accordingly, the author proves some no-go theorems that show, for a broad
class of processes, the order of two events cannot be in a pure superposition,
uncorrelated with any other system. For example this is not possible for a
pure superposition of any pair of Markovian, unitary processes with equal
local dimensions and different causal orders. This result imposes constraints
on novel resources for quantum information processing and on possible
processes in a theory of quantum gravity.

As we see from Fig. (\ref{F3a}) the `before' and the `after' are defined only
relatively. As correctly remarked by Marletto and Vedral \cite{MV}, the only
meaning for the `before' and the `after' is indicated by the order of the
state. This is because the observer does not notice any sense of directivity
as his state only contains information about the instant time. Our remark here
is on constructing a memory; it is hard to understand how the observer can
construct a memory in a universe of stationary states except through the
different\ spatial locations. But if the spatial locations are not known, or
are indistinguishable, then there will be no memory except for the last
location; the one before the transition is made. Accordingly, the treatment of
the transitions from one state to the next could be described using Hidden
Quantum Markov Models \cite{Moras}. In such a treatment the full scope of
transitions would be presented out of the set of possible states of the system
and the corresponding states of the clock. However, this is again an issue
worth further investigation and may provide a simpler description of the
dynamics of the quantum system with a simpler calculation method.

A definite causal order is intimately related to the existence of a pre-set
global time frame independent of the system as it is presented in standard
quantum mechanics. The extra degrees of freedom given for the variables of the
system in a block universe treated within the scope of quantum time presented
briefly in this article, allows for indefinite causal order since in such a
system we have no fixed causal order. On the fundamental level the theorems of
Bell \cite{Bell} and of Kochen and Specker \cite{KS} shows that quantum
mechanics violates the belief that physical observables possess pre-existing
values independent of the context of measurement. But as we have seen above
the causal order in a block universe, as described by relativity theory, is
relative depending on the state of the observer. In the case the internal
quantum time this is adopted the observer inside the universe will see the
projections as per his state of motion relative to the reference before or
after the event as depicted in figure (\ref{F3a}).

For a spatially smeared particle detectors coupled to quantum fields this may
result in breakdown of covariance. This situation is analyzed and evaluated in
\cite{MM} where the predictions of the detectors are considered. However,
although the results of this work may apply to the Unruh-DeWitt model, but
since a perturbative analysis is employed, it is hard to see how it could be
applied to situations where strong gravity is encountered. Furthermore, the
validity of measurements in QFT when performed with non-relativistic particle
detectors is discussed in \cite{Ram}.

\subsection{The Quantum Reference Frames}

The concept of quantum reference frame (QRF) has been originally considered by
Aharonov and Kaufherr \cite{Ahr} who tried to solve the problem of consistency
of the quantum description relative to a finite-mass measuring device in the
non-relativistic case. This they do without appealing to an abstract reference
frame with infinite mass. The proposed solution extends the equivalence
principle to quantum theory where a covariant description relative to a
quantum reference frame is given. However, this solution was offered in
context of a classical description of time in which time is taken as a
parameter, measured with reference to an external fixed frame.

QRFs has also been extensively discussed mainly with the purpose of devising
communication tasks with physical systems serving as detectors
\cite{Bar1,Bar2}. Recently interest revived in considering the QRF by the
introduction of quantum time which requires re-consideration of the question
of the frame of reference, this time in the context of the kinematical
description of quantum states. The necessity for introducing the QRF arises as
we have multiple reference frames. As it is the case in the theory of
relativity a proper time has to be defined to account for the different clocks
which are themselves reference frames and to safeguard the covariance of
measurements \cite{Sina}. Some more work has been done recently on this issue
\cite{Lov,Gia2,Van, Gia}. However, to avoid confusion it is important to note
that some authors may not differentiate explicitly between the \emph{proper
time interval \ }and the \emph{proper time}. In classical relativity we
learned that the proper time $\tau$ is the \textit{wrist watch time}
\cite{Weel}, it is the same for all observers, and it is the reference taken
in the calculations of relativistic physics, whereas the proper time interval
$s$ is different for different observers because it depends on the observer's
world lines. It is obtained when integrating the proper time $\tau$ over the
world line for the particle,%

\begin{equation}
s=\int\limits_{W}d\tau\sqrt{g_{\mu\upsilon}\frac{dW^{\mu}}{d\tau}%
\frac{dW^{\upsilon}}{d\tau}}. \label{qt}%
\end{equation}

In quantum context the problem of the reference frame becomes even more
important in order to resolve the problem of superposition of states. The
clocks at different positions or moving with different speed read different
times. Different observers may read such superpositions of different times of
the different clocks which turns out to be a tensor product of the available
timing of the events.

The Lorentz group has been derived using QRFs from operational conditions on
quantum communication without presupposing a specific spacetime structure
\cite{HM}. This step in dealing with spacetime symmetries is an important move
towards establishing a connection between quantum information and quantum
gravity. Beside this, the metric-independent approach is particularly relevant
in the context of quantum gravity where no fixed notion of spacetime structure
is available. Establishing a "quantum general covariance" is a long-sought
target for an approach to unify quantum mechanics and relativity. Along this
target some steps have already been made. In Ref. \cite{Gia2} an extension of
Galileo's weak equivalence principle for reference frames is developed. The
quantum principle of equivalence suggested by Hardy \cite{Hard} is worth
recognizing, by which it is always possible to transform to a quantum
reference frame and have a definite causal structure in the vicinity of any
given point. Such ideas are indeed worth studying and extending since they
combine several elements of quantum theory and relativity theory in an
endeavor to unify both. These ideas may not form a theory but certainly may be
found helpful when combining some parts or sections of the ultimate theory. As
already been recognized in \cite{Van} the ambition is to compile these
developments through quantum gravity and quantum information into a unifying
method for switching perspectives in the quantum theory that may includes both
spatial and temporal quantum reference systems and applies in both fields.

In another scenario of dealing with problems of unifying general relativity
and quantum mechanics, authors \cite{Ruitz} have recognized that the metric of
general relativity is influenced by matter, and is expected to become
indefinite when matter behaves quantum mechanically. They consider a
generalized PaW approach using several clocks, representing several QRFs.
Accordingly, they develop a framework to operationally define events and their
localization with respect to a quantum clock reference frame in the presence
of gravitating quantum systems. Their results show that the time
localizability of events becomes relative when clocks interact
gravitationally, depending on the reference frame. This relativity identifies
a signature of an indefinite metric, where events can occur in an indefinite
causal order. To deal with such indefiniteness, a reference frame is found
where local quantum operations take their standard unitary dilation form, thus
preserving covariance with respect to quantum reference frame transformations.
The point of interest in this approach is the use of a process matrix
formalism which is a neat presentation of such situations. More recently
Baumann and collaborators \cite{Baum} combined the process matrix framework
with a generalization of the PaW formalism where several agents are
considered, each with their own discrete quantum clock. Process matrices are
then extracted to from scenarios involving such agents with quantum clocks.
The description via a history state with multiple clocks imposes constraints
on the implementation of process matrices and on the perspectives of the
agents as described via causal reference frames. While it allows for scenarios
where different definite causal orders are coherently controlled, the
non-causal processes might not be implemented within this setting. This line
of dealing with quantum clocks and sketching them on\ a landscape of causal
order is a step forward in constructing a temporally ordered metric in the
context of quantum gravity replacing the diffeomorphism of the standard
manifolds of general relativity.

\subsection{Quantum time dilation}

Recently, Cepollaro and collaborators\textit{ }\cite{Cep} investigated the
question whether gravitational time dilation may also be used as a resource in
quantum information theory. They show that the gravitational time dilation may
enhance the precision in estimating gravitational acceleration for long
interferometric times. The crucial result in this work is the affirmation that
interferometric measurements should be performed on both the path and the
clock degrees of freedom. Indeed, this is expected since the proper time
interval depends on the world line, a fact which makes such a result quite plausible.

In this context comes the problem of time dilation in quantum clocks. One
facet of this problem appears once we know that clocks are ultimately quantum
systems, as any other quantum system they are subject to the superposition
principle too. Unlike the classical case, in a relativistic context, this
leads to the possibility of the quantum clocks experiencing a superposition of
proper times. Such scenarios have been investigated in the context of
relativistic clock interferometry. According to general relativity also,
proper time flows at different rates in different regions of spacetime.
Several articles has been recently published to investigates this topic. One
of the early articles considering a quantum effect that cannot be explained
without the general relativistic notion of proper time was by Zych and
collaborators \cite{Zych}\ where they consider interference of a `clock'
represented by a particle with evolving internal degrees of freedom that will
not only display a phase shift, but also reduce the visibility of the
interference pattern. Because of quantum complementarity, the visibility will
drop to the extent to which the path information becomes available from
reading out the proper time from the clock. Therefore, such a gravitationally
induced decoherence could provide a test of the genuine general relativistic
notion of proper time in quantum mechanics. In fact, this is one of the
preliminary investigations which lead to the proposal of Bose-Marletto-Vedral
(BMV) experiment \cite{Bose,Mar} for testing a quantum gravitational effect
predicted by low energy purturbative quantum gravity. If detected such a test
would provide indirect empirical evidence that spacetime geometry obeys
quantum mechanics.

Smith and Ahmadi \cite{SA} considered the question of time dilation in quantum
clocks, where they introduce a proper-time observable defined as a covariant
POVM on the internal degrees of freedom of a relativistic particle moving
through curved spacetime. This allowed considering two relativistic quantum
clocks $A$ and $B$ for which they construct the probability that $A$ reads a
particular proper time conditioned on $B$ reading a different proper time.
They compute this probability distribution under such a condition by extending
the PaW approach. A more elegant analysis we find in Ref. \cite{Gia} where the
author considers a system of $N$ particles in a weak gravitational field
trying to introduce a timeless formulation along the approach of PaW to obtain
a frozen state of the $N$ particles. Such frozen states are featured in the
Block universe formalism of PaW. Then the dynamics of such a system is
regained by the relational dynamics. This is obtained by having the observer
inside the closed system or the block universe. This approach clarifies many
other presentations of the same problem.

\section{Discussion and Conclusions}

Nearly a decade ago Moreva and collaborators \cite{Mor} suggested a
well-elaborated experiment to verify the PaW proposal on quantum measurement
of time taking into consideration the modifications invoked by Gambini and
collaborators \cite{Gam}. The experiment uses two photons in a static
entangled state one of them being a clock which uses the polarization states
$H$ and $V$ as reference for time, and the other being the evolving system. It
was shown in \cite{Mor} that an internal observer will see the system
developing in the time of the internal clock while an external
(super-observer) will see the system static, hence verifying the picture of
block universe implied by the PaW proposal. Our quantum time analogy with the
reel of a movie depicted in Fig (\ref{F3a}) perfectly agrees with the
description given in this experiment.

Gravity is the lab were quantum features of time can be exposed. Strong
gravitational fields found in the vicinity of neutron stars and black holes is
the best venue to look for exposition of such features. In such regions
quantum vacuum and gravity rules the interplay of an emergent dynamical time.
It is important to note that the concept of quantum time involves two basic
requirement: the first is that time is a relatival entity that takes its
meaning through the relational dynamic of change and, the second, is that time
can be measured with reference to an internal clock correlated with the system
to mark the occurrence of events. Realizing this connection between quantum
time and the main character of quantum information as a confined system of
causal structure is a key element for innovative developments in quantum
information processing. In fact, encoding quantum information through
relational degrees of freedom is expected to simplify the calculations since
in some cases we can use hidden quantum Markov models (HQMM) and consequently
adapt the calculation into simpler models \cite{Moras,Clark,Jav}. The beauty
of these models is that the conditional probability is not restricted to have
the state occurring at a given time necessarily but that the state is
occurring after or before a designated state. This makes the dynamics truly
relational. Beside this, it has recently been shown that quantum clocks are
more accurate than the classical ones as quantum clocks can achieve a
quadratically improved accuracy compared to purely classical clock of the same
size \cite{Woods}. Combined with the conditional probability argument in the
PaW approach this result implies a large accuracy in defining the energy
states avoiding large portion of the inherent uncertainty in the energy of the
system when a classical clock for measuring time is adopted.

Perhaps there is now enough justification to abandon the background-dependent
approaches to quantum gravity, but it is hard to see how one can neglect time,
for without time we will face a problem with the order of events. Causal
relationships will be ambiguous without order of time. Furthermore, the
conventional thermodynamic arrow of time may be lost although we may be able
to replace it with a quantum arrow of time which takes into consideration the
increase of entanglement as a measure for the direction of time as suggested
by Marletto and Vedral \cite{MV}. It seems that this is proposed as an
alternative to the measure of entropy, as it is known that if the whole
universe is in a pure state, its entropy is always zero.

Taken in a very broad scope of relationships, the dynamics of any system is an
expression of the causal effects from within the available variables of the
system. Classical and quantum observables are those variables that we can
directly measure. This explains the role played by the causal structure and
the identified causal order in exposing the dynamics of the system. The fact
that indefinite causal order allows for more degrees of freedom within the
same set of variables and, on the same footing, having quantum entanglement
and coherence giving rise to quantum-enhanced information processing, the
power of quantum computation with indefinite causal structures may lead to new
protocols and procedures that may even change the character of quantum
information itself \cite{Pearl}. However, we also agree with \cite{Buck} in
saying that \textquotedblleft the present research programme will not reach
fulfilment if it does not provide new insights into the challenge of finding a
theory of quantum gravity.\textquotedblright

Starting from an overall quantum description of two entangled but
non-interacting systems, one of which is counted as a clock, Foti and
collaborators \cite{Foti} take the classical limit of the clock and obtains
the Schr\"{o}dinger equation in this limit. Upon taking the classical limit
for both the clock and the evolving system, they obtain Hamilton's equation of
motion. In their opinion, this shows that there is not a \textquotedblleft
quantum time\textquotedblright which is possibly opposed to a
\textquotedblleft classical\textquotedblright\ one; there is only one time and
it is a manifestation of quantum entanglement. These results can be easily
explained by knowing that the time in Schr\"{o}dinger's equation is continuous
and consequently recovering the equation in the classical limit is expected.
The Hamiltonian of the global system is formed of the clock Hamiltonian and
the system Hamiltonian, and upon taking the classical limit of both systems,
the clock Hamiltonian gets dissolved and we are left with Hamilton's equation
of motion. Indeed, there are now two types of time, a discrete quantized time
and a classical continuous time.

It might be of importance here to notice that states in the block universe are
stationary when observed from a hypothetical external observer, despite being
under time development with respect to the internal observer. This status of
states being stationary as seen by a hypothetical external observers might be
utilized in a technique that could be prepared in the light of the work of
Stobinska et. al., \cite{Stob} to enhance the information processing speed.
Here I am expecting that such utilization can be performed in within the
information processing package.

In conclusion we\ may say that at this stage of the development of quantum
time it seems that the formulation of a consistent scheme for this important
topic which is necessary for the progress in quantum gravity research and
quantum information development, has already reached an advanced level
overcoming many fundamental conceptual and technical difficulties. But this is
taking place at the price of changing the paradigm. The formulations presented
above are consistent and beautiful. What remains is the next step of employing
temporal relational dynamic to play its role on a temporal causal structure
that may become a viable replacement of the spacetime manifolds of the theory
of general relativity, exposing quantum features like non-locality merged with
the fundamental features of relativity in a simple and elegant
picture.\bigskip

\begin{acknowledgments}
The author would like to thank Almut Beige, for support and encouragement to
write this article, Daniel Hodgson for help in preparing some figures, Vlatko
Vedral, Lorenzo Maccone, Tommaso Favalli, Alexander Smith, Flaminia Giacomini,
Philip H\"{o}hn and Alessandrro Coppo for useful discussions during the
preparation of this work
\end{acknowledgments}

\end{document}